\documentclass[aps,pre,superscriptaddress,amsmath,amssymb,twocolumn,nofootinbib]{revtex4}

\usepackage[dvipdfmx]{graphicx}

\usepackage{bm}
\usepackage{color}
\usepackage{amsmath}
\usepackage{here}

\def\Vect#1{\mbox{\boldmath $#1$}}
\def\phic{\phi_{\rm iso}}
\def\phisj{\phi_{\rm SJ}}
\def\phith{\phi_{\rm th}}
\def\gadst{\gamma_{\rm DST}^{\rm (I)}}

\begin{document}

\title{
  Shear jamming, discontinuous shear thickening, and fragile states in dry granular materials under oscillatory shear
}

\author{Michio Otsuki}
\email[]{otsuki@me.es.osaka-u.ac.jp}
\affiliation{
  Graduate School of Engineering Science, Osaka University, Toyonaka, Osaka 560-8531, Japan}
\author{Hisao Hayakawa }
\affiliation{Yukawa Institute for Theoretical Physics, Kyoto University, Kitashirakawaoiwake-cho, Sakyo-ku, Kyoto 606-8502, Japan}



\begin{abstract}
We numerically study the linear response of two-dimensional frictional granular materials 
under oscillatory shear.
The storage modulus $G'$ 
and the loss modulus $G''$ 
in the zero strain rate limit depend on the initial strain 
amplitude of the oscillatory shear
before measurement.
The shear jammed state (satisfying $G'>0$) 
can be observed at an amplitude greater than a critical initial strain amplitude.
The fragile state is 
defined by the emergence of liquid-like and solid-like states
depending on the form of the initial shear. In this state, the
observed $G'$ after the reduction of
the strain amplitude depends on the phase of the external shear strain.
The loss modulus $G''$
exhibits a discontinuous jump
corresponding to discontinuous
shear thickening in the fragile state. 

\end{abstract}
\date{\today}


\maketitle

\section{Introduction}

Amorphous disordered materials such as
granular media, colloidal suspensions, foams, and emulsions
interact via dissipative and repulsive forces, and 
in a dense regime these systems can form solid-like jammed states.
Liu and Nagel proposed 
a phenomenological phase diagram for jamming transition,
in which a particulate system jams upon compression,
however, it unjams upon application of a shear force \cite{Liu}.
This proposal attracted much attention
among physicists \cite{Hecke,Behringer18}. 
The existence of jammed states have 
been verified 
in several numerical simulations of frictionless granular particles.
In these works, a continuous change of pressure as well as a
discontinuous change of the coordination number has been
observed across the jamming transition density \cite{OHern02,OHern03,Wyart05}.
Other researchers have reported various critical scaling laws 
of rheological quantities near the jamming density
for frictionless particles 
up to the present time
under steady shear
\cite{Olsson, Hatano07, Hatano08, Tighe, Hatano10, Otsuki08, Otsuki09, Otsuki10, Nordstrom, Olsson11, Vagberg, Otsuki12, Ikeda, Olsson12, DeGiuli, Vagberg16, Boyer, Trulsson, Andreotti, Lerner, Vagberg14, Kawasaki15, Suzuki, Rahbari}
and oscillatory shear \cite{Tighe11,Otsuki14}.


In reality due to the roughness of particles,
mutual friction between particles
is unavoidable in granular systems.
Bi et al. \cite{Bi11} suggested that the jamming process qualitatively 
differs between frictional and frictionless grains; in frictional systems, 
shear forces counterintuitively induces jammed states 
even below the friction-dependent critical fraction $\phic$.
Such transition, known as shear jamming, has been extensively studied both
experimentally \cite{Zhang08,Zhang10,Wang18,Zhao} 
and numerically \cite{Sarkar13,Sarkar16,Seto19,Pradipto}.
Bi et al. \cite{Bi11} further proposed
the existence of a fragile state in a system under pure shear
characterized by
the percolation of the force chain only in the compressive direction
\cite{Footnote1}.
In contrast, the force chain in a shear jammed state percolates in all the directions.
However, the definition of the fragile state in Ref. \cite{Bi11} is non-quantitative
and inapplicable to other systems, necessitating
a quantitative definition.

The mutual friction between granular particles results
in a distinct rheological transition 
known as discontinuous shear thickening (DST)
\cite{Otsuki11,Chialvo,Brown,Seto,Fernandez,Heussinger,Bandi,Ciamarra,Mari,Grob,Kawasaki14,Wyart14,Grob16,Hayakawa16,Hayakawa17,Peters,Fall,Sarkar,Singh,Kawasaki18,Thomas}.
DST is important in industrial applications such as 
protective vests,
robotic manipulators,
and traction controls \cite{Brown14,Brown10}.
Several papers have investigated
the relation 
between DST and shear jamming
in suspensions of frictional grains
under steady shear \cite{Peters,Fall,Sarkar,Singh}.
In stress-controlled experiments, DST can be observed over a wide region of the phase diagram
\cite{Peters}; however, in rate-controlled experiments, 
DST can be observed only as a boundary line between shear jamming and continuous shear thickening
in the phase diagram \cite{Fall}.
Because these results seem to be inconsistent, the relation between 
shear jamming and DST is not yet clarified.

To resolve the aforementioned problems,
we numerically 
measure the complex shear modulus in two-dimensional frictional grains 
near the jamming point under oscillatory shear.
Therefore, we apply
the discrete element method (DEM) \cite{Cundall}.
In Sec. \ref{Model},
we explain our setup and model.
Section \ref{G:sec} deals 
with effects of initial oscillatory shear 
on the shear modulus.
In Sec. \ref{Phase:sec},
we clarify the relations 
among the shear jammed state, the fragile state, 
and the DST-like behavior by controlling 
the initial strain amplitude $\gamma_0^{\rm (I)}$
and the area fraction $\phi$.
We discuss and conclude our results in Sec. \ref{Discussion}.
In Appendix \ref{Model:app},
we explain the details of our simulation.
In Appendix \ref{mu}, we discuss the dependence of transition points
on the friction coefficient $\mu$.
The dependence of the phase diagram 
on the number of the oscillatory shear is discussed in Appendix \ref{Nc}.
In Appendix \ref{Sec:s},
we explain how the shear jammed state
appears in the stress-strain curve.
In Appendix \ref{Rs},
we show the fabric
anisotropy of the contact network
in our simulation.

\section{Setup of our simulation}
\label{Model}

Let us consider a two-dimensional assembly of $N$ frictional granular particles
having the identical density $\rho$ confined in a square box of linear size $L$.
The inter-particle interactions are modeled as linear springs with
the normal and tangential spring constants of $k^{\rm (n)}$ and $k^{\rm (t)}$, respectively, the friction coefficient $\mu$, and the restitution coefficient $e$
\cite{Cundall}.
DEM is detailed in Appendix \ref{Model:app}.
To avoid crystallization, we construct a bidispersed
system with
an equal number of grains of two diameters ($d_0$
and $d_0/1.4$).
We also adopt $N=4000$, $k^{\rm (n)}=0.2k^{\rm (t)}$,
$\mu = 1.0$, and $e = 0.043$.

\begin{figure}[htbp]
\includegraphics[width=1.0\linewidth]{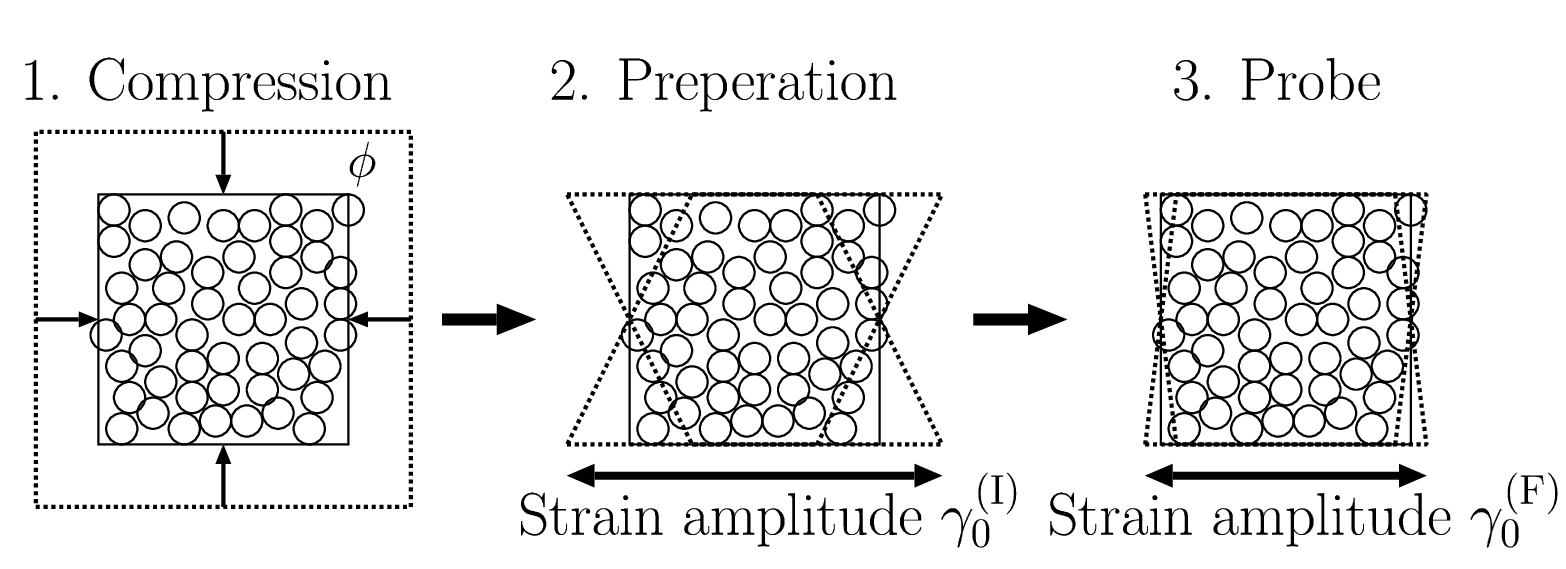}
\caption{
  Illustration of the protocol.
}
\label{Process}
\end{figure}

At the beginning of our simulation, the frictional disks are randomly placed
with the area fraction $\phi_{\rm I}=0.75$,
and we slowly compress the system
until the area fraction reaches a designated value $\phi$
as shown in Fig. \ref{Process}. 
In each step of the compression process, we increase 
the area fraction by
$\Delta \phi = 10^{-4}$
with the affine transformation,
and relax grains to a mechanical equilibrium state
where the kinetic temperature $T < T_{\rm th}=10^{-8} k^{\rm (n)}d_0^2$.
We have confirmed that the shear modulus after the compression
is insensitive to the choices of $T_{\rm th}$ and $\Delta \phi$
if $T_{\rm th} \le 10^{-8} k^{\rm (n)}d_0^2$
and 
$\Delta \phi \le 10^{-4}$.
Note that we estimate the isotropic jamming point 
$\phic=0.821$ for $\mu=1.0$, which might depend on the preparation procedure \cite{Luding,Kumar}.
See Appendix \ref{mu} for the determination and $\mu$-dependence of $\phic$.

\begin{figure}[htbp]
  \begin{center}
    \begin{tabular}{c}
      \begin{minipage}{0.5\hsize}
      \begin{center}
	\includegraphics[width=1.0\linewidth]{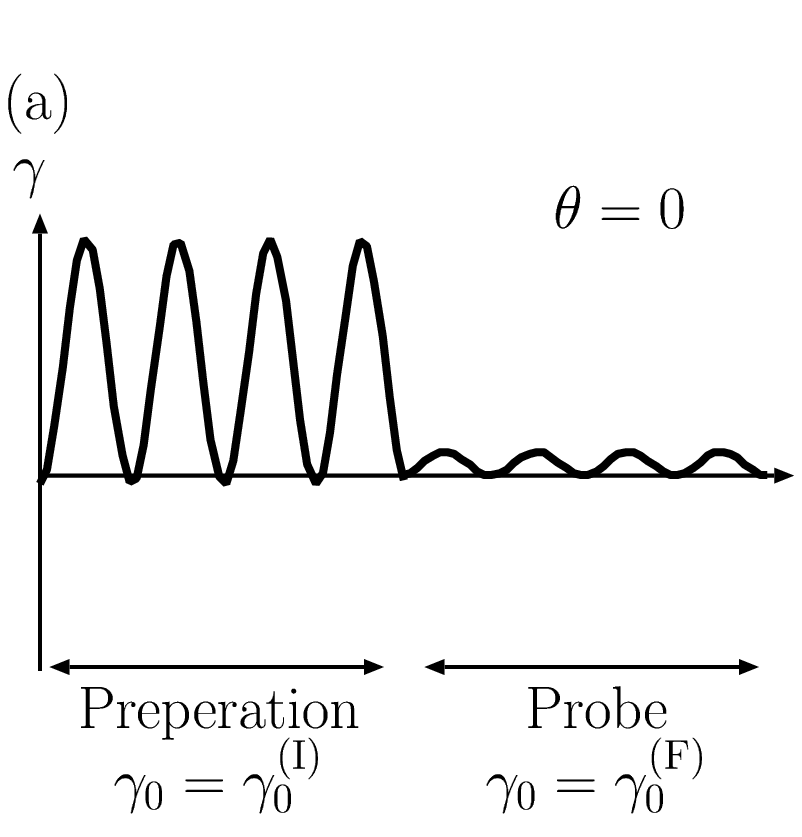}
      \end{center}
      \end{minipage}
      \begin{minipage}{0.5\hsize}
      \begin{center}
	\includegraphics[width=1.0\linewidth]{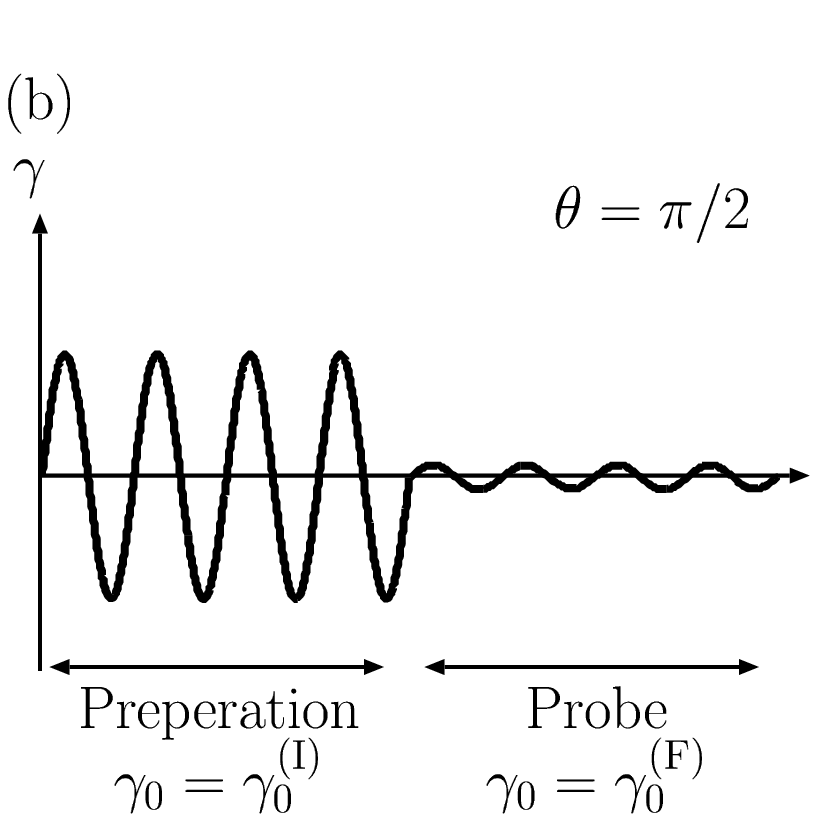}
      \end{center}
      \end{minipage}
    \end{tabular}
\caption{
  Plots of the shear strain $\gamma(t)$ against $t$
  for $\theta=0$ (a) and $\pi/2$ (b).
}
\label{Strain}
  \end{center}
\end{figure}

We further apply the shear strain 
\begin{equation}
\gamma (t) = \gamma_0 
\left \{ \cos\theta - \cos(\omega t + \theta) \right \}
\end{equation}
in the $x$-direction
of the compressed system using the SLLOD method \cite{Evans}
as shown in Fig. \ref{Process}.
Here, $\gamma_0$, $\omega$, and $\theta$
denote the strain amplitude, the angular frequency,
and the initial phase, respectively.
We apply $N_c^{\rm (I)}$ cycles of
the oscillatory shear with
the initial strain amplitude $\gamma_0 = \gamma_0^{\rm (I)}$
as a preparation of the system.
Note that $\theta$ controls the asymmetricity
of the applied strain
as shown in Fig.\ \ref{Strain},
where the strain-free state 
($\gamma(t)=0$) is the minimum strain state for $\theta = 0$,
while it lies between the maximum and minimum values
for $\theta = \pi/2$.
After the cycles, 
we reduce the strain amplitude
to $\gamma_0 = \gamma_0^{\rm (F)}=1.0 \times 10^{-4}$
and apply $N_c^{\rm (F)}$ cycles
of the oscillatory shear as a probe to
measure the storage modulus $G'$ and the loss modulus $G''$
in the linear response region. Here,
$G'$ and $G''$ are, respectively,
defined by \cite{Doi} 
\begin{eqnarray}
  G' & = & - \frac{\omega}{\pi } \int_{0}^{2 \pi/\omega} \ dt \
\sigma(t) \cos(\omega t + \theta)/\gamma_0^{\rm (F)}, \\
  G'' & = & \frac{\omega}{\pi} \int_{0}^{2 \pi/\omega} \ dt \
\sigma(t) \sin(\omega t + \theta)/\gamma_0^{\rm (F)}.
\end{eqnarray}
The moduli $G'$ and $G''$ are measured in the final cycle.
The shear stress $\sigma$ in the above expressions is
 given by
\begin{eqnarray}
  \sigma = - \frac{1}{2L^2} \sum_{i} \sum_{j>i}
  \left ( r_{ij,x} F_{ij,y} + r_{ij,y} F_{ij,x} \right ), 
  \label{sigma}
\end{eqnarray}
where $F_{ij,\alpha}$ and $r_{ij,\alpha}$ denote the $\alpha$ components of 
the interaction force $\Vect{F}_{ij}$
and the relative position vector $\Vect{r}_{ij}$
between grains $i$ and $j$, respectively.
The contributions of
the kinetic part of $\sigma$ 
and the coupled stress 
(i.e., the asymmetric part of the shear stress)
are ignored because they are less
than $1 \%$ of $\sigma$.
Note that 
$G'$ and the dynamic viscosity $\eta(\omega) \equiv G''(\omega)/\omega$ 
corresponding to the apparent viscosity
 are almost independent of $\omega$
and $\gamma_0^{\rm (F)}$
when $\omega \le 10^{-2} t_0^{-1}$, $\gamma_0^{\rm (I)} \le 1.0$, and
 $\gamma_0^{\rm (F)}\le 10^{-3}$
 with
$t_0 = \sqrt{m_0/k^{\rm (n)}}$ and the mass $m_0$ 
for a grain with the diameter $d_0$ \cite{Otsuki17}. 
Thus, we investigate
only the effects of $\gamma_0^{\rm (I)}$, $\theta$, and $\phi$ on the shear modulus, fixing 
$\omega = 10^{-4} t_0^{-1}$
and $\gamma_0^{\rm (F)} = 10^{-4}$.
We have also confirmed that $G'$ and the phase diagram of the system
are almost 
independent of $N_c^{\rm (I)}$ and $N_c^{\rm (F)}$
when $N_c^{\rm (I)}\ge 10$ and $N_c^{\rm (F)}\ge10$ as shown in Appendix 
\ref{Nc}
and used $N_c^{\rm (I)} = N_c^{\rm (F)} = 10$.
We adopt the leapfrog algorithm
with the time step $\Delta t = 0.05 t_0$.

\section{Mechanical response} 
\label{G:sec}

\begin{figure}[htbp]
  \begin{center}
    \begin{tabular}{c}
      \begin{minipage}{0.33\hsize}
      \begin{center}
	\includegraphics[width=1.0\linewidth]{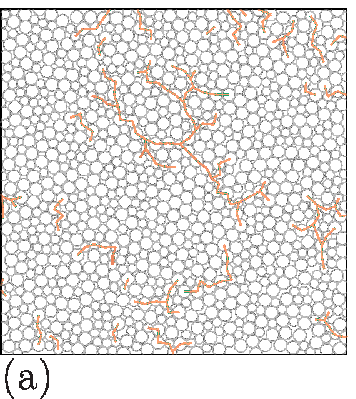}
      \end{center}
      \end{minipage}
      \begin{minipage}{0.33\hsize}
      \begin{center}
	\includegraphics[width=1.0\linewidth]{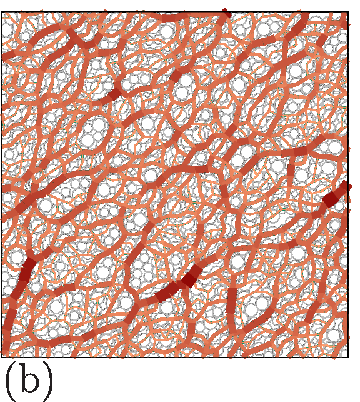}
      \end{center}
      \end{minipage}
      \begin{minipage}{0.33\hsize}
      \begin{center}
	\includegraphics[width=1.0\linewidth]{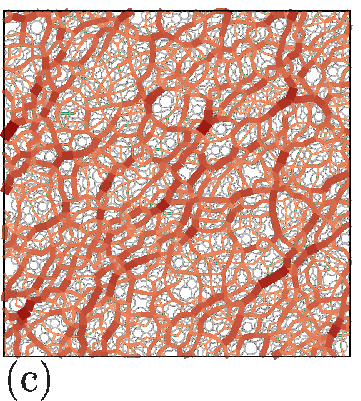}
      \end{center}
      \end{minipage}
    \end{tabular}
\caption{
The snapshots of grains (circles)
and force chains (lines) for $\phi=0.820$ and $\theta = 0$
immediately after 
the strain amplitudes
(a) $\gamma_0^{\rm (I)} = 0.1$, (b) $0.12$, and (c) $1.0$ are reduced to $\gamma_0^{(F)}=1.0\times 10^{-4}$. Panels (a), (b), and (c)
correspond to the unjammed, fragile, and shear jammed
states, respectively.
The color and width of each line depend on the absolute
value of the interaction force between grains.
}
\label{F_SJ}
  \end{center}
\end{figure}

Figure \ref{F_SJ}
displays the force chains
immediately after the reduction of the strain amplitude
for $\phi=0.820<\phic$, and $\theta = 0$
with $\gamma_0^{\rm (I)} = 0.1, 0.12$, and $1.0$.
When the initial strain amplitude is small ($\gamma_0^{\rm (I)}=0.1$),
the system remains in a liquid-like state
with no percolating force chains.
Under high initial strains ($\gamma_0^{\rm (I)} = 0.12$ and $1.0$), 
the system develops anisotropic percolating force chains.
Unlike the expectation in Ref. \cite{Bi11}, the shear jammed state seems to have anisotropic percolating force chains.

\begin{figure}[htbp]
\includegraphics[width=1.0\linewidth]{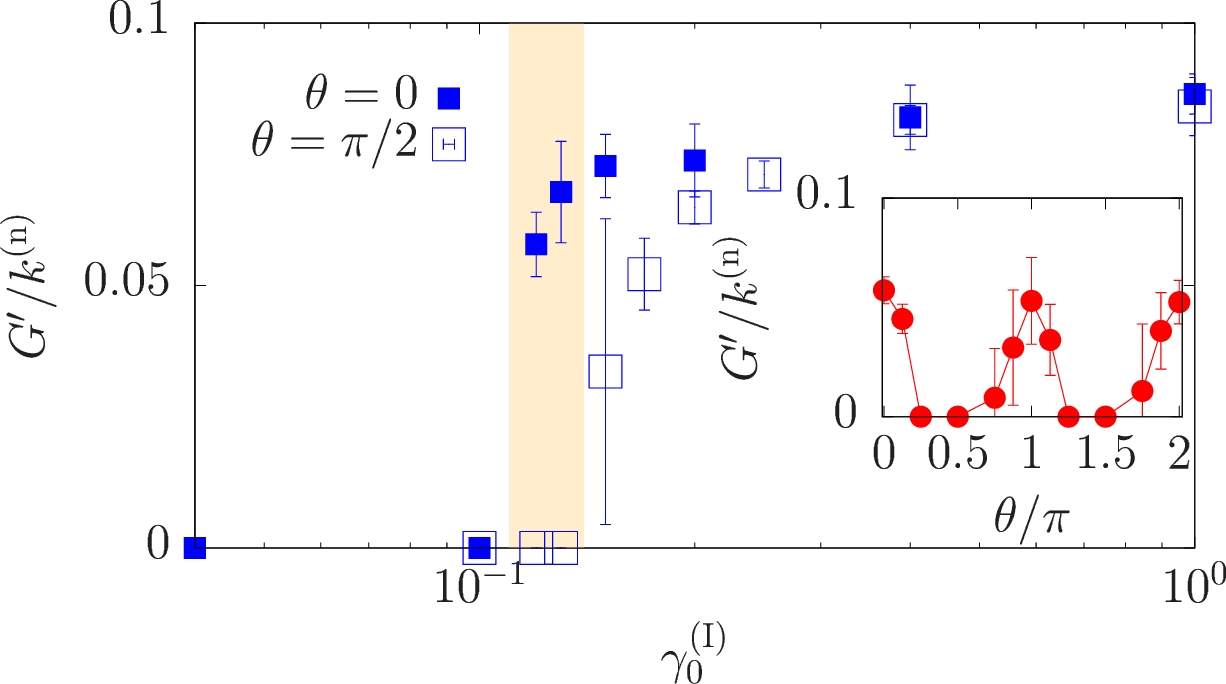}
\caption{
Plots of the storage modulus $G'$ versus 
$\gamma_0^{\rm (I)}$ for $\phi=0.82$
with $\theta = 0$ and $\pi/2$.
The shaded region highlights the fragile state.
Inset: 
Storage modulus $G'$ versus
$\theta$ for $\phi=0.82$
with $\gamma_0^{\rm (I)}=0.12.$
}
\label{G_ga_theta}
\end{figure}

Figure \ref{G_ga_theta} displays $G'$ versus $\gamma_0^{\rm (I)}$
for $\theta=0$ and $\pi/2$ with $\phi=0.820$. The shear induces transitions
from a liquid-like to a solid-like state.
See Appendix \ref{Sec:s} for the shear induced jamming
in the stress-strain curve of the initial oscillation.
$G'$ strongly depends on $\theta$
near the critical strain amplitudes
(shaded region of Fig. \ref{G_ga_theta}).
The inset of Fig. \ref{G_ga_theta}
displays $G'$ versus 
$\theta$ for $\phi=0.82$
and $\gamma_0^{\rm (I)}=0.12$.
The storage modulus $G'$ peaks at $n \pi$
and falls to $0$ near $(n + 1/2) \pi$,
where $n$ is an integer.

\begin{figure}[htbp]
\includegraphics[width=1.0\linewidth]{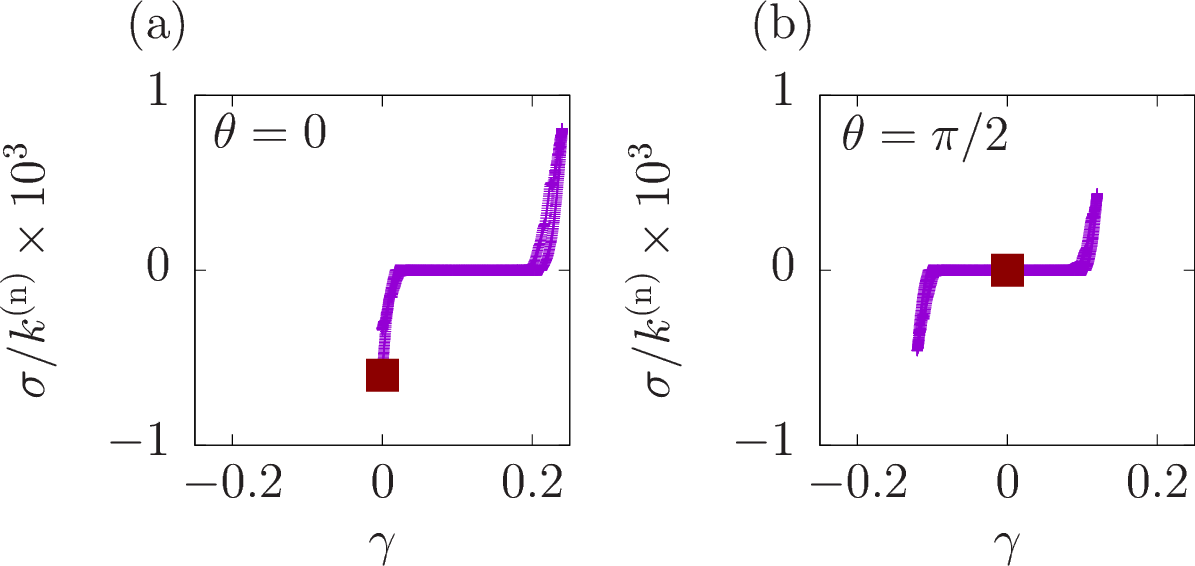}
\caption{
Plots of the shear stress $\sigma$ versus the strain $\gamma$
in the last cycle of the initial oscillatory shear
with $\gamma_0^{\rm (I)}=0.12$
 and $\phi=0.820$ at $\theta = 0$ (a) and $\pi/2$ (b).
 The solid squares indicate
 the positions of the linear response measurements after the strain amplitude is reduced.
}
\label{s_ga_nu0.82000}
\end{figure}

Figure \ref{s_ga_nu0.82000}
displays the shear stress $\sigma(t)$ versus the strain $\gamma(t)$
in the last cycle of the initial oscillation with $\gamma_0^{\rm (I)}=1.2$
and $\phi=0.820$ at $\theta = 0$ and $\pi/2$.
When $\theta=0$, the shear stress $\sigma$ can be fitted by a linear function of the strain $\gamma$ 
near the maximum and minimum values of $\sigma$, but remains $0$ over $0.03<\gamma<0.2$
(Fig. \ref{s_ga_nu0.82000} (a)).
The linear response near $\gamma \approx 0$
in the stress-strain curve
is consistent with the solid-like state
after the reduction of the strain amplitude (i.e., $G'>0$ at $\theta=0$).
Setting $\theta = \pi/2$ shifts the stress--strain curve of the initial oscillation without significantly changing its shape
from that of $\theta = 0$
(see Fig. \ref{s_ga_nu0.82000}(b)).
In this case, the linear response near $\gamma\approx 0$
denotes the liquid-like state
after the reduction of the strain amplitude (i.e., $G'=0$).
These results explain the $\theta$-dependence of $G'$
in Fig. \ref{G_ga_theta}.

\begin{figure}[htbp]
\includegraphics[width=1.0\linewidth]{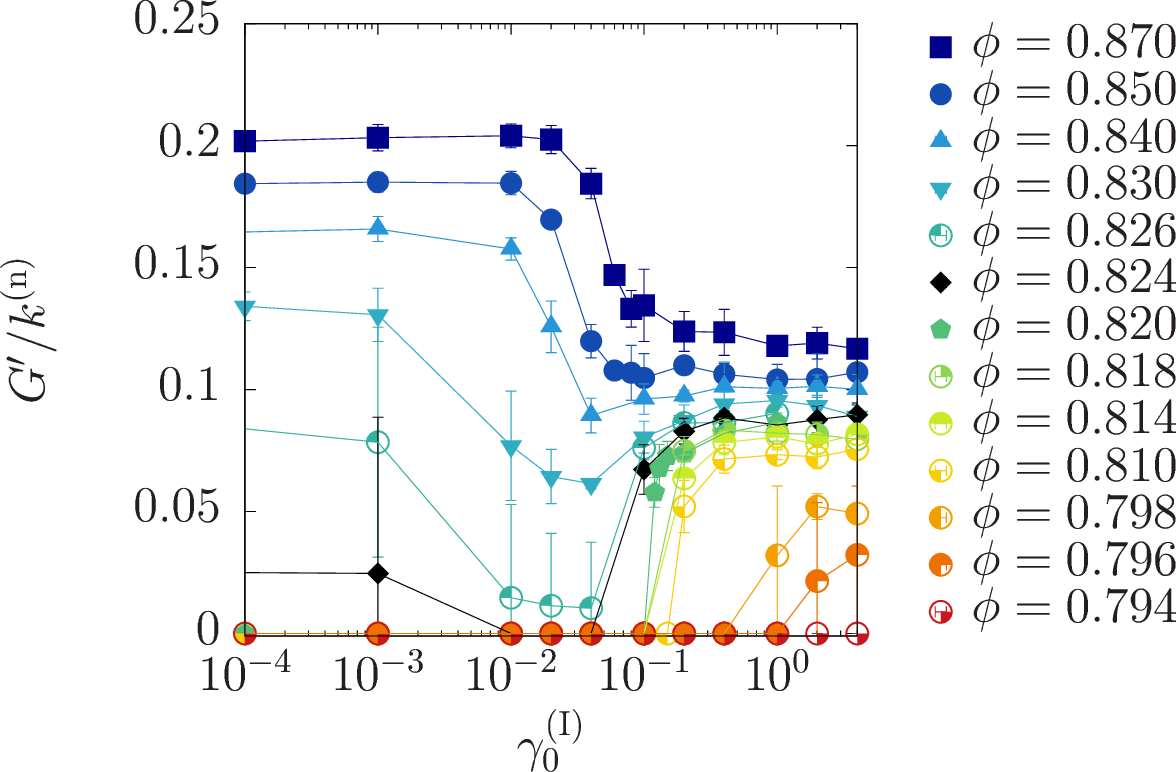}
\caption{
  Plots of the storage modulus $G'$ versus 
$\gamma_0^{\rm (I)}$ for various $\phi$
at $\theta = 0$.
}
\label{G_ga_MC}
\end{figure}

Figure \ref{G_ga_MC}
displays the storage modulus $G'$ versus
$\gamma_0^{\rm (I)}$ for various $\phi$
at $\theta=0$.
When $\phi>\phic=0.821$, $G'$ is finite for $\gamma_0^{\rm (I)}=0$,
but depends on $\gamma_0^{\rm (I)}$.
When $\phi>0.84$, $G'$ is a decreasing function of
$\gamma_0^{\rm (I)}$, which corresponds to the softening observed in glassy materials under steady-shear conditions \cite{Fan}.
In $0.82 < \phi < 0.84$, $G'$ is minimized at intermediate values of $\gamma_0^{(I)}$.
Shear jamming is observed in $\phisj <\phi < \phic$,
where $\phisj=0.795$ (as determined in Appendix \ref{mu}).
We also observe a re-entrant behavior at $\phi=0.824$,
where $G'$ changes from $G'>0$ to $G' \simeq 0$ and reverts to $G'>0$ at higher $\gamma_0^{(I)}$.

\begin{figure}[htbp]
\includegraphics[width=1.0\linewidth]{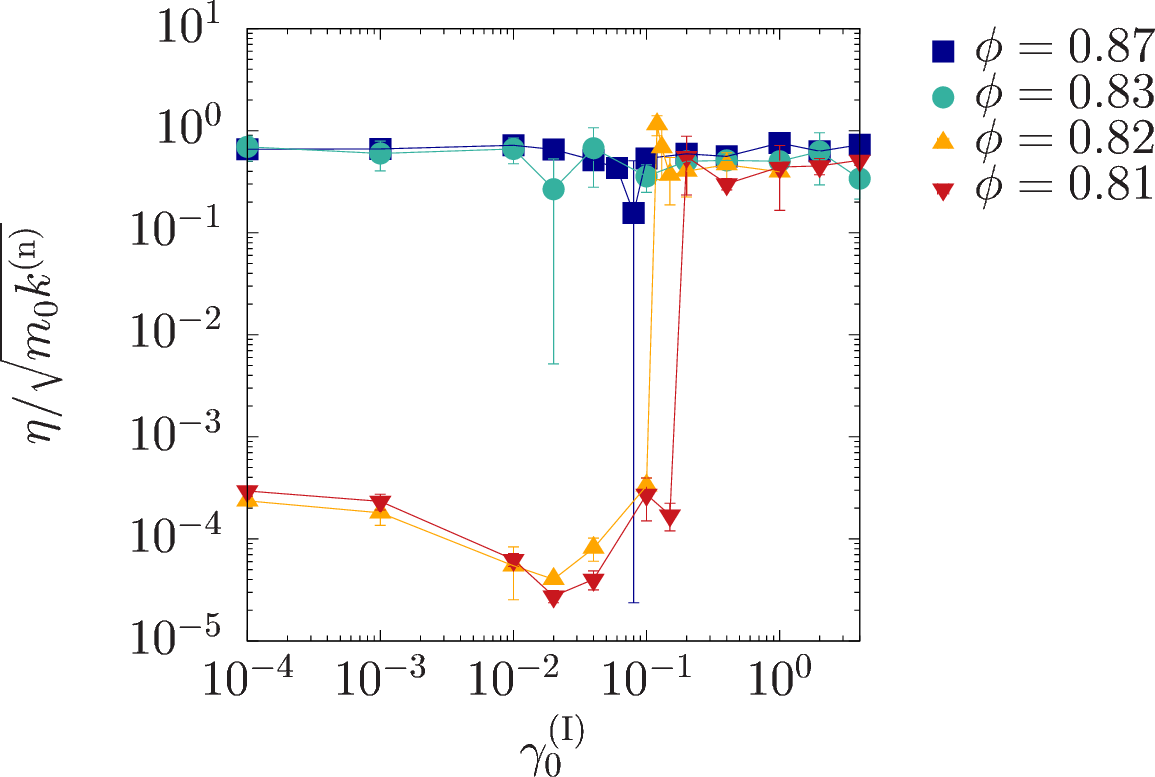}
\caption{
Plots of the dynamic viscosity $\eta$ 
versus the initial strain amplitude $\gamma_0^{\rm (I)}$
for various $\phi$ at $\theta=0$.}
\label{G2_ga}
\end{figure}

Figure \ref{G2_ga} displays the dimensionless dynamic viscosity
 versus $\gamma_0^{\rm (I)}$
 for $\theta=0$ and various $\phi$.
The viscosity $\eta$ is almost independent
of $\gamma_0^{\rm (I)}$ when $\phi$ exceeds $\phic$,
but jumps from a negligibly small value
to a large value in $\phisj < \phi < \phic$. This discontinuity, which takes place
at a critical amplitude of the initial strain $\gadst$,
corresponds to DST under steady shear.

\section{Phase diagram}
\label{Phase:sec}
\begin{figure}[htbp]
  \includegraphics[width=0.7\linewidth]{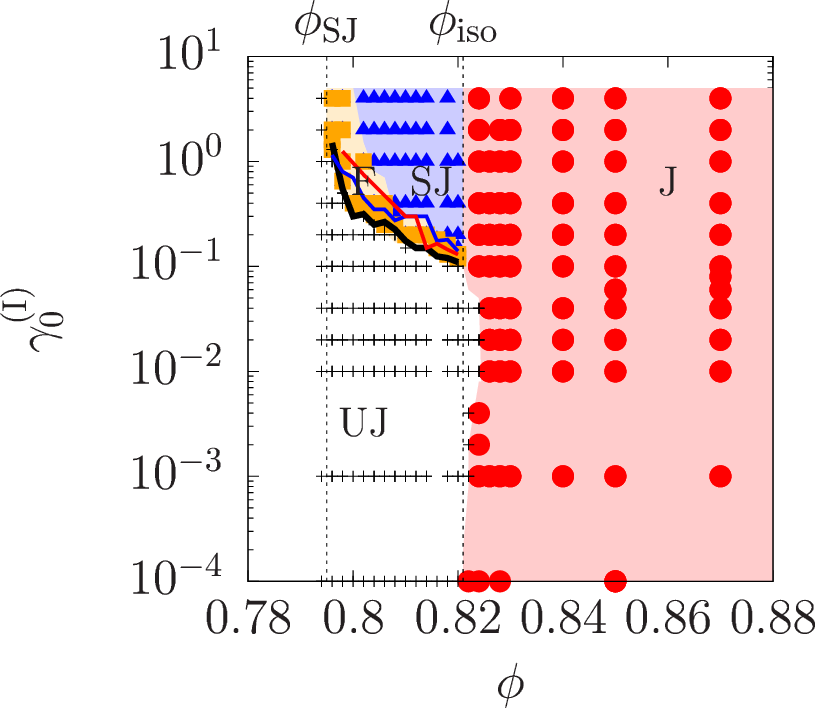}
\caption{
Phase diagram 
 on the $\phi$ versus $\gamma_0^{\rm (I)}$ plane. 
Circles, triangles, squares, and crosses
represent the J, SJ, F, and UJ states, respectively.
The thick black line, thin blue line, and thin red line represent
the critical strain amplitudes $\gadst$ at $\theta=0$, $\pi/4$, and $\pi/2$,
respectively.
}
\label{Phase_ga_MC}
\end{figure}

Figure \ref{Phase_ga_MC} depicts
the phase diagram on the $\gamma_0^{\rm (I)}$ versus $\phi$ plane.
Here, we have introduced the shear storage modulus without initial oscillatory shear as
$G'_0(\phi) \equiv \lim_{\gamma_0^{\rm (I)} \to 0} G'\left(\phi,\gamma_0^{\rm (I)}\right)$.
We then define the jammed (J) state 
in which $G'_0(\phi)>G_{\rm th}$ 
and
$G'\left(\phi, \gamma_0^{\rm (I)}\right) > G_{\rm th}$
for any $\theta$
with a sufficiently small threshold $G_{\rm th} = 10^{-4} k^{\rm (n)}$.
Note that the phase diagram
is unchanged by setting $G_{\rm th} = 10^{-5} k^{\rm (n)}$.
The unjammed (UJ) state is defined as
 $G'\left(\phi, \gamma_0^{\rm (I)}\right) < G_{\rm th}$
for any $\theta$, and the shear jammed (SJ) state is defined as
 $G'_0(\phi)<G_{\rm th}$ and
$G'\left(\phi, \gamma_0^{\rm (I)}\right) > G_{\rm th}$
for any $\theta$.
 Finally, 
 in the fragile (F) state,
 whether the state is solid-like with
 $G'\left(\phi, \gamma_0^{\rm (I)}\right) > G_{\rm th}$ 
or liquid-like with $G'\left(\phi, \gamma_0^{\rm (I)}\right) < G_{\rm th}$
depends on the value of $\theta$
(see the inset of Fig. \ref{G_ga_theta}).
 In Fig. \ref{Phase_ga_MC}, the SJ state
 exists in the range $\phisj < \phi < \phic$
 and $\gamma_0^{\rm(I)} > 0.1$.
 Remarkably, the UJ phase exists even when $\phi>\phic$, 
 and the J state at large $\gamma_0^{(I)}$ 
 and $\phi>\phic$ (located above the bay-like unjammed state),
 which is observed in a numerical simulation of frictionless particles under oscillatory shear \cite{Das}.
 This may be regarded as an SJ-like state. However, this state differs from 
 the SJ state defined as the memory effect of the initial strain as introduced above.
 We have also confirmed the existence of the fragile state between the UJ and SJ states.

Figure \ref{Phase_ga_MC}
 also displays the critical strain amplitude
$\gadst$ for the DST-like behavior,
where the viscosity $\eta$ exceeds a threshold $10^{-3} \sqrt{m_0 k^{\rm (n)}}$.
 Note that at $\gadst$, $G'$ simultaneously changes from $0$
 to a finite value.
 When $\theta$ is $0$, the critical strain amplitude $\gadst$ resides on the boundary between
 the UJ and fragile states, whereas at other $\theta$, it
 resides in the fragile state.
 This suggests that the fragile state
 exhibits the DST-like behavior 
 at least when $\gamma_0^{(I)}$ is not excessively large.

 \section{Discussion and concluding remarks}
 \label{Discussion}
 Let us now discuss our results.
 Recent numerical simulations
 with different protocols
 indicated that shear jamming
 occurs even in frictionless systems
\cite{Kumar,Jin,Urbani,Jin18,Bertrand,Baity,Chen18, VinuthaN,VinuthaJ,VinuthaA}.
 In our simulation, the SJ state disappears
 at $\mu=0$ (see Appendix \ref{mu}).
 Nevertheless, 
 the re-entrant process
 in the range $\phic<\phi<0.826$ of our system
 seems to be related with 
 the SJ states in frictionless systems.

\begin{figure}[htbp]
\includegraphics[width=0.8\linewidth]{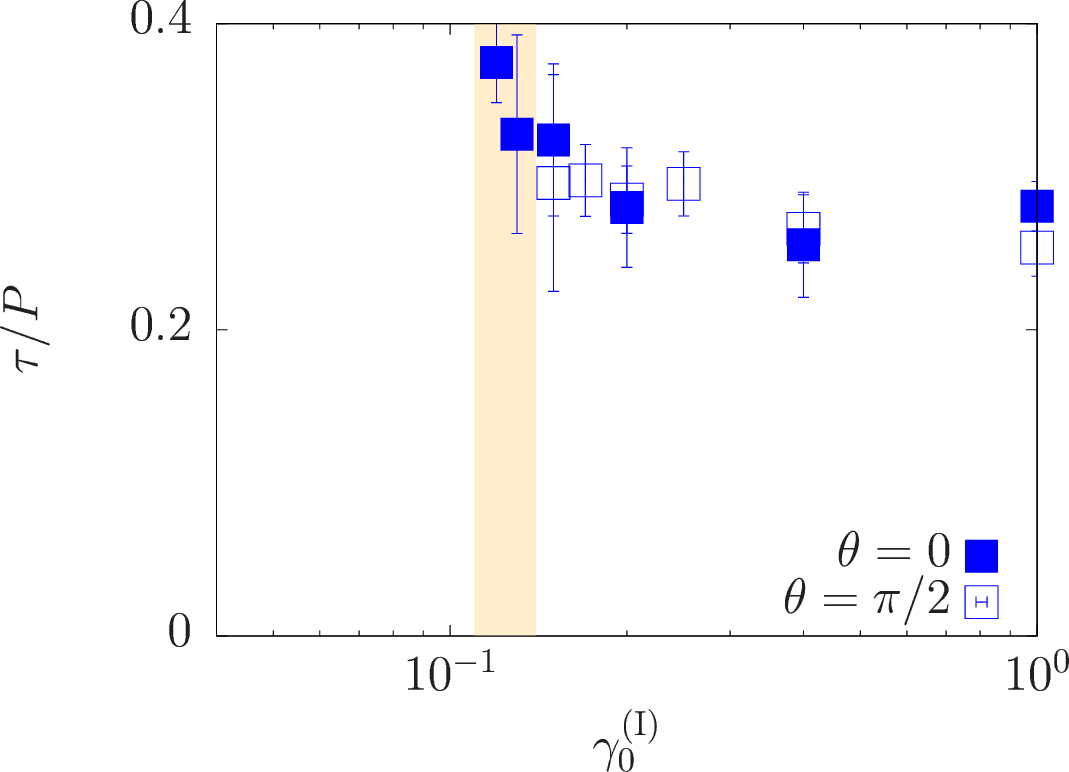}
\caption{
Plots of the stress anisotropy $\tau/P$
 versus 
 $\gamma_0^{\rm (I)}$ for $\phi=0.820$
 with $\theta = 0$ and $\pi/2$.
 The shaded region highlights the fragile state.
}
\label{tau_P_ga_theta}
\end{figure}

 The fragile state was originally defined by the anisotropic percolation
 of force chains under a quasi-static pure shear process
\cite{Bi11}.
 Because
 the compressive direction 
 changes with time and no quasi-static operations are imposed in our system, we cannot apply the original argument based on percolation networks 
(Fig. \ref{F_SJ}(b)).
 Regardless, the stress anisotropy 
 $\tau/P$ \cite{Sarkar16,Thomas,Chen18} immediately after 
the reduction of the initial strain amplitudes
is maximized in the fragile state and remains constant 
in the SJ state as shown in Fig. \ref{tau_P_ga_theta}. In this figure,
$\tau=(\sigma_1-\sigma_2)/2$
 and $P=-(\sigma_1+\sigma_2)/2$, where
$\sigma_1$ and $\sigma_2$ denote the maximum and minimum principal stresses,
 respectively.
 This behavior is qualitatively similar to that 
 for 
 the experimentally observed behavior \cite{Sarkar16}
 and 
   the fabric anisotropy shown in Appendix \ref{Rs}.
 It is possibly explained by a phenomenology based on
 the probability distribution of 
 sliding forces \cite{DeGiuli17}.
The mutual relation between the fragile state and the anisotropy
requires further careful investigation.

 In conclusion, 
 we have numerically studied frictional granular systems
 under oscillatory shearing.
 By controlling the strain amplitude $\gamma_0^{\rm (I)}$
 of the oscillatory shear
 before measurement,
 we have observed that 
a solid-like state with the storage modulus $G'>0$
is induced.
This indicates that 
shear jamming is an effect of the preparation (cyclic shear), 
which could be seen as memory effect.
 We have also observed a fragile state 
 in which 
 the linear response can be solid-like with $G'>0$ or liquid-like with
 $G'=0$ depending on the phase of the oscillation.
 This protocol has also detected
 the DST-like behavior, 
 manifesting a remarkable discontinuity in the viscosity versus the initial strain plot.
 The region of the DST-like behavior in the phase diagram is 
 almost identical with that of the fragile state.

\begin{acknowledgments}
The authors thank R. Behringer, B. Chakraborty, 
T. Kawasaki, C. Maloney, C. S. O'Hern,  K. Saitoh, S. Sastry, S. Takada, and
H. A. Vinutha for fruitful discussions. 
We would like to dedicate this paper to the memory of R. Behringer,
who has passed away in July, 2018.
This work is partially supported by the Grant-in-Aid of MEXT for Scientific 
Research (Grant No. 16H04025, No. 17H05420, and No. 19K03670).
One of the authors (M.O.) appreciates the warm hospitality of Yukawa Institute for Theoretical Physics at Kyoto University during his stay there supported by the Program No. YITP-T-18-03 and YITP-W-18-17.
One of the authors (HH) thanks the warm hospitality of ICTS.
\end{acknowledgments}

\appendix

\section{Details of our DEM}
\label{Model:app}

In this appendix, we present the details of our DEM.
To suppress shear bands, we apply an
oscillatory shear along the $x$-direction under Lees--Edwards boundary conditions using the SLLOD method \cite{Evans}.
The equation of motion of grain $i$
(the mass $m_i$, the position $\bm{r}_i=(x_i,y_i)$,
and the diameter $d_i$) is written as
\begin{equation}
  m_i \frac{d^2}{dt^2}{\bm{r}}_i=\bm{F}_{i}.
  \label{S1}
\end{equation}
The total force $\bm{F}_{i}$ acting on the grain is given by
\begin{eqnarray}
  \bm{F}_{i} & = & \sum_{j \neq i} 
  \left (F_{ij}^{\rm (n)}\Vect{n}_{ij} + F_{ij}^{\rm (t)}\Vect{t}_{ij} \right ) 
  \nonumber \\
  & = & 
\sum_{j \neq i} 
  \left(
    \begin{array}{cc}
      \cos \alpha_{ij} &  -\sin \alpha_{ij} \\
      \sin \alpha_{ij}  & \cos \alpha_{ij} \\
    \end{array}
  \right)
  \left(
    \begin{array}{c}
      F_{ij}^{\rm (n)}\\
      F_{ij}^{\rm (t)}\\
    \end{array}
  \right)
\end{eqnarray}
with the normal contact force $F_{ij}^{\rm (n)}$,
the tangential contact force $F_{ij}^{\rm (t)}$,
the normal unit vector $\Vect{n}_{ij}$, and
the tangential unit vector $\Vect{t}_{ij}$
between grains $i$ and $j$.
$\Vect{n}_{ij}$ and
$\Vect{t}_{ij}$ respectively satisfy
$\Vect{n}_{ij}=(\cos \alpha_{ij},\sin \alpha_{ij})$ and
$\Vect{t}_{ij}=(-\sin \alpha_{ij},\cos \alpha_{ij})$
with $\alpha_{ij} = \tan^{-1}((y_i-y_j)/(x_i-x_j))$.
The normal contact force $F_{ij}^{\rm (n)}$ is given by $F_{ij}^{\rm (n)}
= -\left ( k^{\rm (n)} u_{ij}^{\rm (n)}
+ \zeta^{\rm (n)} v_{ij}^{\rm (n)} \right )
\Theta(d_{ij}-r_{ij})$
with the normal displacement $u_{ij}^{\rm (n)} = r_{ij}-d_{ij}$,
$d_{ij}=(d_i+d_j)/2$, $r_{ij}=|\bm{r}_{ij}|=|\bm{r}_i-\bm{r}_j|$,
the normal velocity 
$v_{ij}^{\rm (n)} = ({\Vect{v}}_i - {\Vect{v}}_i)
\cdot \Vect{n}_{ij}$,
the velocity ${\Vect{v}}_i$ of grain $i$,
the normal spring constant $k^{\rm (n)}$,
and the normal damping constant $\zeta^{\rm (n)}$.
$\Theta(x)$ is the Heviside step function satisfying 
$\Theta(x)=1$ for $x\ge 0$ and $\Theta(x)=0$ otherwise.
The tangential force is given by $F_{ij}^{\rm (t)} =
{\rm min}\left(  |\tilde F_{ij}^{\rm (t)}|,\mu F_{ij}^{\rm (n, el)}\right )
  {\rm sgn}\left(  \tilde F_{ij}^{\rm (t)} \right )
\Theta(d_{ij}-r_{ij})$,
    where $\mu$ is the friction coefficient, ${\rm min}(a,b)$ selects the smaller one
    between $a$ and $b$,
    ${\rm sgn}(x)=1$ for $x\ge 0$ and ${\rm sgn}(x)=-1$ otherwise, 
    and $\tilde F_{ij}^{\rm (t)}$ is given by
    $\tilde F_{ij}^{\rm (t)} =  -k^{\rm (t)} u_{ij}^{\rm (t)} - 
    \zeta^{\rm (t)} v_{ij}^{\rm (t)}$
with the tangential spring constant $k^{\rm (t)}$ and 
the tangential damping constant $\zeta^{\rm (t)}$.
The tangential velocity 
$v_{ij}^{\rm (t)}$ 
and the tangential displacement
$u_{ij}^{\rm (t)}$, respectively, 
satisfy
$v_{ij}^{\rm (t)}
=  ({\Vect{v}}_i - {\Vect{v}}_i)
\cdot \Vect{t}_{ij} - (d_i \omega_i+d_j \omega_j)/2$
and
$\dot{u}_{ij}^{\rm (t)}=v_{ij}^{\rm (t)}$
for $|\tilde F_{ij}^{\rm (t)}| < \mu F_{ij}^{\rm (n, el)}$
with the angular velocity $\omega_i$ of grain $i$.
If $|\tilde F_{ij}^{\rm (t)}| \ge \mu F_{ij}^{\rm (n, el)}$,
$u_{ij}^{\rm (t)}$ remains unchanged.
We note that $u_{ij}^{\rm (t)}$ is zero 
if grains $i$ and $j$ are detached.

We adopt $N=4000$, $\mu=1.0$,
$k^{\rm (t)} = 0.2 k^{\rm (n)}$, and
$\zeta^{\rm (t)} = \zeta^{\rm (n)}= \sqrt{m_0 k^{\rm (n)}}$
in this paper.
This set of parameters corresponds
to the constant restitution coefficient 
\begin{equation}
  e = \exp \left ( - \frac{\pi}{\sqrt{2 k^{\rm (n)}m_0/\zeta^{\rm (n)} - 1 }}\right ) \simeq 0.043
\end{equation}
for a grain with the diameter $d_0$.

\section{Determination of transition points and their dependence on $\mu$}
\label{mu}

In this appendix, we first
explain how to determine $\phic$ 
for the jamming and $\phisj$ for the shear jamming.
We also
discuss the $\mu$-dependence of these transition points.

\begin{figure}[htbp]
\includegraphics[width=0.7\linewidth]{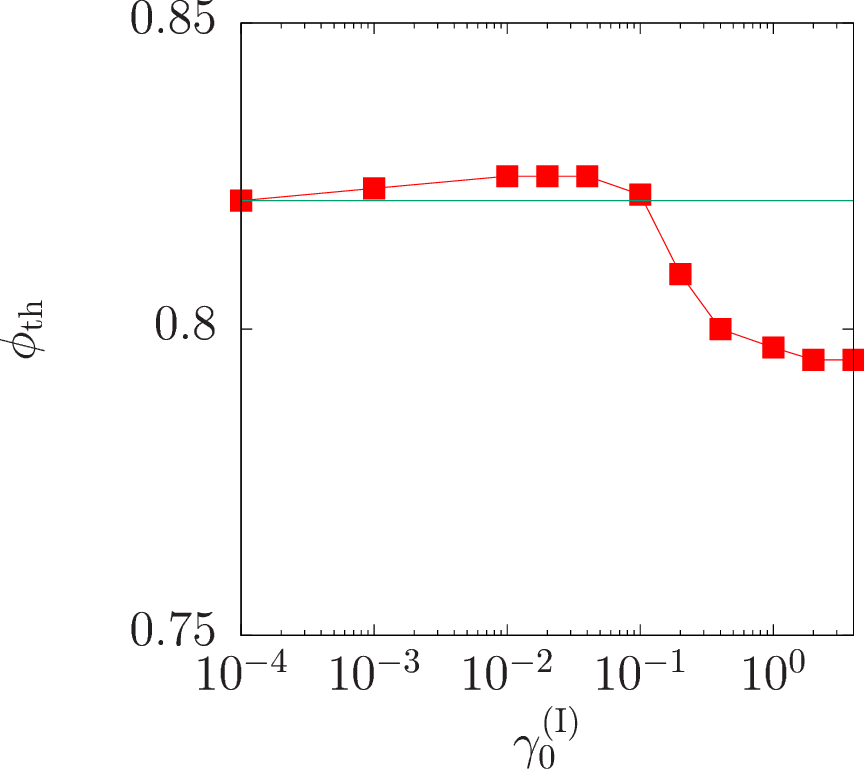}
\caption{
  Plots of the transition point $\phith$ versus $\gamma_0^{\rm (I)}$
  for $\theta=0$ and $\mu=1.0$.
  The solid thin line 
  parallel to the horizontal axis
  represents $\phic$.
}
\label{phic_ga}
\end{figure}

For a given set of $\gamma_0^{\rm (I)}$ and $\theta$,
the storage modulus $G'$ 
exhibits a transition from $G'=0$ to $G'>0$
at a transition point $\phith(\gamma_0^{\rm (I)},\theta)$.
In Fig. \ref{phic_ga},
we plot the transition point $\phith(\gamma_0^{\rm (I)},\theta)$
  versus $\gamma_0^{\rm (I)}$
  for $\theta=0$ and $\mu=1.0$.
The transition point increases 
with $\gamma_0^{(I)}$ for $\gamma_0^{(I)}<0.04$,
and decreases with $\gamma_0^{(I)}$ for $ \gamma_0^{(I)} > 0.04$.
This dependence of the transition point on the preparation
is consistent with the concept of the moving jamming point
used to explain the shear jamming in Ref. \cite{Kumar}.
Then, we define the jamming point
without shear as 
  \begin{equation}
  \phic \equiv \lim_{\gamma_0^{\rm (I)}\to0}\phith(\gamma_0^{\rm (I)},\theta),
  \end{equation}
  which is independent of $\theta$ by definition.
  It should be noted that $\phic$ is the isotropic jamming point
  under sufficiently small compression rate
  without any overcompression used in Ref. \cite{Kumar}.
We also define 
the transition point for the shear jamming as
  \begin{equation}
\phisj \equiv \min_{\gamma_0^{\rm (I)}, \theta} \phith(\gamma_0^{\rm (I)},\theta).
\label{phisj}
\end{equation}
Within our observation, $\phith(\gamma_0^{(I)},\theta)$ takes its smallest value at $\theta=0$ and seems to converge for sufficiently large $\gamma_0^{(I)}$.
We, thus, evaluate
$\phisj$ as $\phith(\gamma_0^{\rm (I)}=4.0,\theta=0)$,
 which is the transition point at the largest initial strain amplitude we apply in our simulation.

\begin{figure}[htbp]
\includegraphics[width=0.7\linewidth]{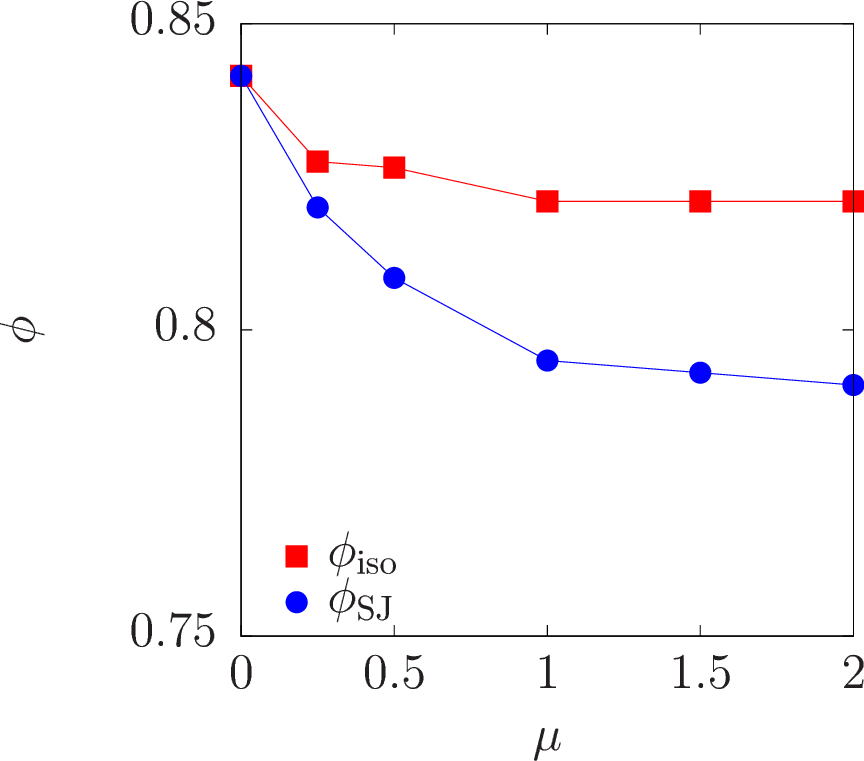}
\caption{
Plots of the transition points $\phic$ and $\phisj$ versus $\mu$.
}
\label{phic}
\end{figure}

In the main text, we have presented the data only for $\mu=1.0$, but
we show the $\mu$-dependence of the critical points $\phic$ and
$\phisj$ in Fig. \ref{phic}.
Note that the shear jamming in terms of Eq. \eqref{phisj} 
is observed only for $\phisj \le \phi \le \phic$.
As shown in Fig. \ref{phic},
the difference between $\phic$ and $\phisj$ decreases
as $\mu$ decreases.
Then,
the shear jamming
based on our definition
disappears
in the frictionless limit.
This, however, does not deny the shear jamming in frictionless grains
in different protocols.
Indeed,
shear jamming in frictionless grains has been reported in previous studies
\cite{Kumar,Jin,Urbani,Jin18,Bertrand,Baity,Chen18,
 VinuthaN,VinuthaJ,VinuthaA}.

\section{The dependence of the phase boundaries on $N_c^{\rm (I)}$}
\label{Nc}

\begin{figure}[htbp]
\includegraphics[width=1.0\linewidth]{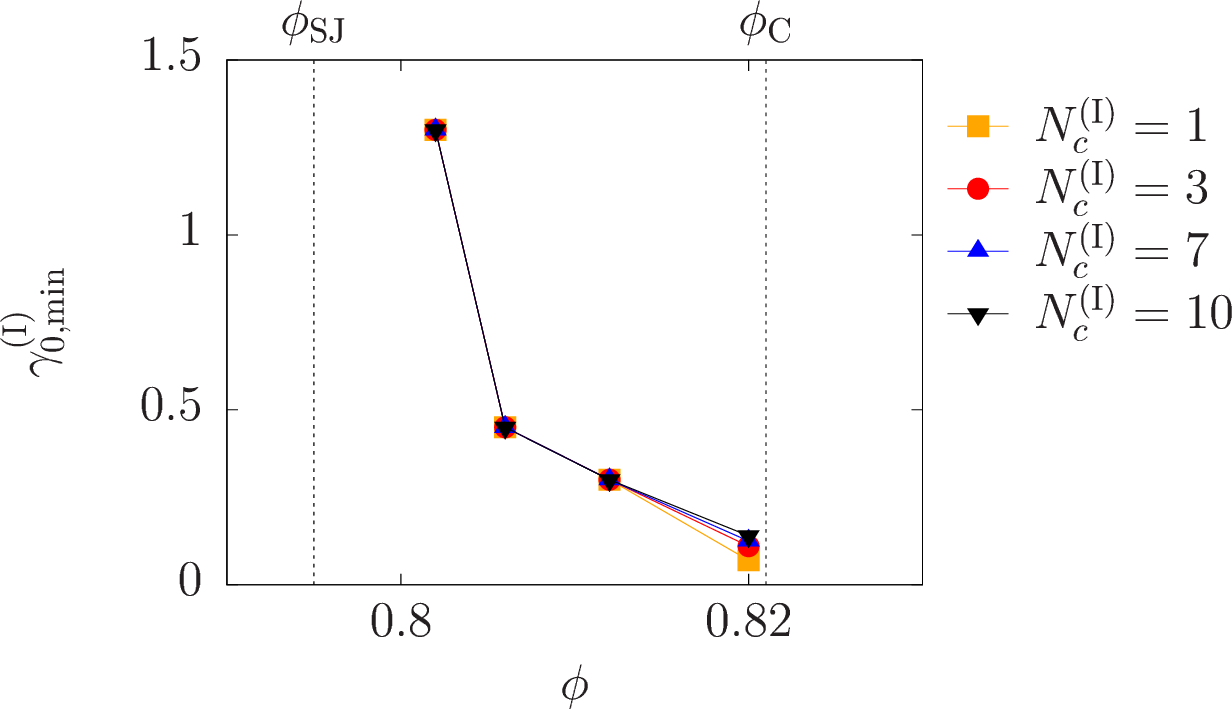}
\caption{
 Plots of $\gamma_{0,{\rm min}}^{\rm (I)}(\phi)$ 
 versus $\phi$ for various $N_c^{\rm (I)}$.
}
\label{gc}
\end{figure}

In this section, we show the dependence of the phase diagram
on the number $N_c^{\rm (I)}$ of cycles in the initial oscillatory shear.
Here, we introduce the minimum strain amplitude
$\gamma_{0,{\rm min}}^{\rm (I)}(\phi)$ for SJ,
where $G'(\phi,\gamma_0^{\rm (I)})>G_{\rm th}$ for any $\theta$
if $\gamma_0^{\rm (I)} > \gamma_{0,{\rm min}}^{\rm (I)}(\phi)$.
It should be noted that $\gamma_{0,{\rm min}}^{\rm (I)}(\phi)$
gives the boundary between the SJ and F states in Fig. \ref{Phase_ga_MC}.
In Fig. \ref{gc}, we plot $\gamma_{0,{\rm min}}^{\rm (I)}(\phi)$ 
versus $\phi$ for various $N_c^{\rm (I)}$, where
$\gamma_{0,{\rm min}}^{\rm (I)}(\phi=0.82)$
slightly increase with $N_c^{\rm (I)}$, 
though $\gamma_{0,{\rm min}}^{\rm (I)}(\phi)$ 
is insensitive to $N_c^{\rm (I)}$ for $\phi \le 0.81$.
Therefore, we safely state that 
$\gamma_{0,{\rm min}}^{\rm (I)}(\phi)$
converges
for $N_c^{\rm (I)} \ge 10$
and arbitrary $\phi$.

\section{Initial stress-strain curve and the shear jamming}
\label{Sec:s}

In this appendix, 
we explain how the shear jamming 
in the linear response regime is related to
the initial stress-strain curve for large strain amplitudes. 
We also explain the reason why the liquid-like response 
can be observed if the initial strain amplitude is sufficiently small.

\begin{figure}[htbp]
\includegraphics[width=0.8\linewidth]{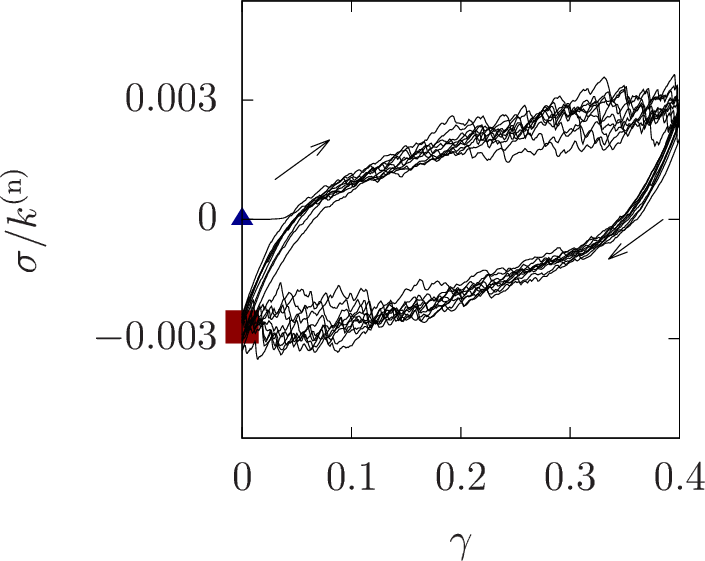}
\caption{
Plots of the shear stress $\sigma$ versus the strain $\gamma$
for $\gamma_0^{\rm (I)}=0.2$,
$\phi=0.820$, and $\theta = 0$.
The triangle and the square indicate 
the states
before and after the initial oscillatory shear,
respectively.
The arrows 
indicate the direction of time evolution in the stress-strain curve.
}
\label{s_ga}
\end{figure}

In Fig. \ref{s_ga}, we plot
the shear stress $\sigma$ versus the strain $\gamma$
for $\gamma_0^{\rm (I)}=0.2$,
$\phi=0.820$, and $\theta=0$.
  Note that $\gamma_0^{\rm (I)}=0.2$
  for this area fraction
corresponds to the shear jammed state.
The stress $\sigma$ follows a stress-strain loop 
once $\gamma$ exceeds $\gamma \simeq 0.02$.
Even after the reduction of the strain amplitude,
there is finite gradient of $\sigma$ against $\gamma$
around $\gamma=0$
which is equivalent to $G'>0$.
Note that the red filled square in Fig. \ref{s_ga} is the measurement point.
This emergence of $G'>0$ is regarded as the occurrence of the shear jamming.

Figure \ref{s_ga} is useful to understand the reason why we observe the liquid-like response if $\gamma_0^{(I)}$ is small for $\phi=0.82$ and $\theta=0$.
Indeed, $\sigma$ remains almost zero for $\gamma\le 0.01$ in this figure. Then, if we reduce $\gamma_0$ to $\gamma_0^{(F)}=1.0\times 10^{-4}$, we only obtain $G'=0$ for $\gamma_0^{(I)}\le 0.01$.

\section{Fabric anisotropy of contact network}
\label{Rs}

\begin{figure}[htbp]
\includegraphics[width=0.9\linewidth]{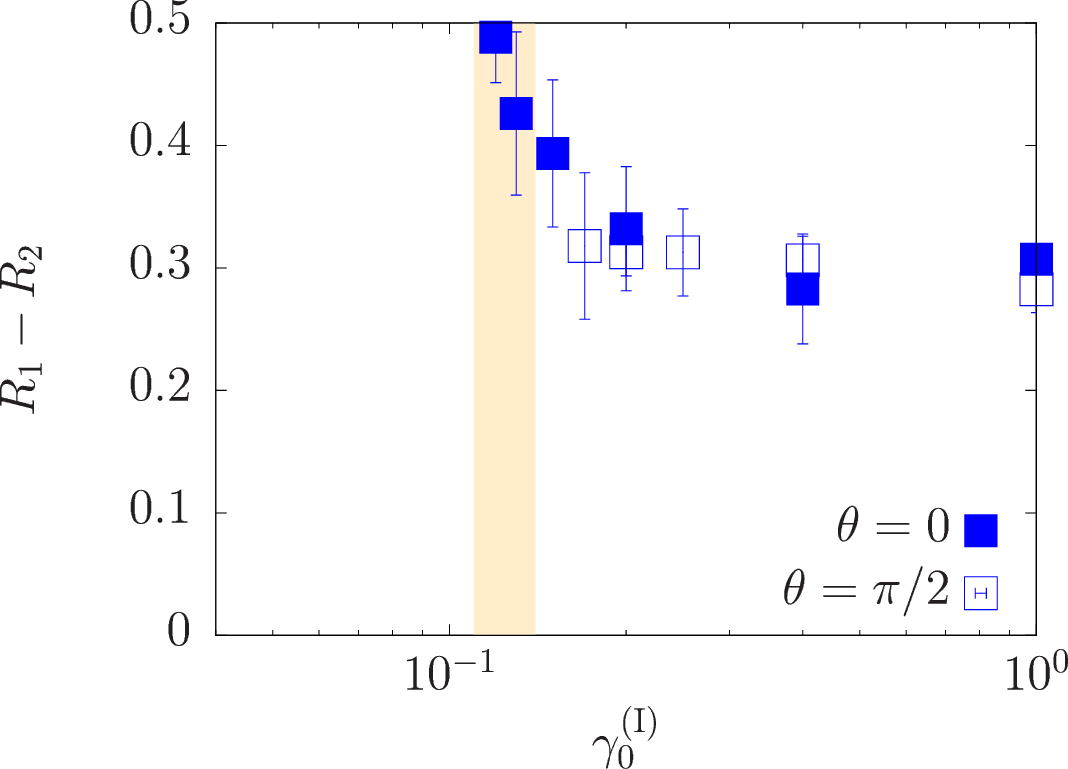}
\caption{
  Plots of the fabric anisotropy $R_1 - R_2$
  versus $\gamma_0^{\rm (I)}$ for $\phi=0.820$
  with $\theta=0$ and $\pi/2$.
  The shaded region corresponds to the fragile state.
}
\label{rho}
\end{figure}

In this appendix, we present the result of the fabric anisotropy of the contact network
in the fragile and shear jammed states.
Let us introduce the contact fabric tensor $R_{\alpha \beta}$ 
as \cite{Bi11}
\begin{eqnarray}
  R_{\alpha \beta} = \frac{1}{N} \sum_{i} \sum_{j>i}
  \frac{r_{ij,\alpha} r_{ij,\beta}}{r_{ij}^2}\Theta(d_{ij} - r_{ij}).
  \label{R}
\end{eqnarray}
Figure \ref{rho} displays
the fabric anisotropy $R_1 - R_2$
  versus $\gamma_0^{\rm (I)}$ for $\phi=0.820$
  with $\theta=0$ and $\pi/2$,
  where the maximum and the minimum eigenvalues 
  of $R_{\alpha \beta}$ are denoted as $R_1$ and $R_2$, respectively.
The fabric anisotropy takes the maximum
in the fragile state and keeps constant in SJ,
which corresponds to the stress anisotropy
in Fig. \ref{tau_P_ga_theta}.

\end{document}